\documentclass[aps,twocolumn,showpacs]{revtex4}
\usepackage{amsmath}
\usepackage{amsfonts}
\usepackage{amssymb}
\usepackage[dvips]{graphicx}
\usepackage[linktocpage,breaklinks=true,colorlinks=true,linkcolor=black,filecolor=black,citecolor=black,pagecolor=black,urlcolor=black]{hyperref}

\newcommand{\cf}{cf.\ }
\newcommand{\eg}{e.\,g.\ }
\newcommand{\ie}{i.\,e.\ }

\newcommand{\dd}{\mathrm{d}}
\newcommand{\vp}{\operatorname{vp}}
\newcommand{\realt}{\operatorname{Re}}
\newcommand{\ii}{\mathrm{i}}
\newcommand{\ew}[1]{\left\langle{#1}\right\rangle}

\newcommand{\DC}{\Delta_\mathrm{c}}
\newcommand{\omrec}{\omega_\mathrm{R}}
\newcommand{\Erec}{E_\mathrm{R}}
\newcommand{\kB}{k_\mathrm{B}}
\newcommand{\vT}{v_\mathrm{T}}
\newcommand{\etac}{\eta_\mathrm{c}}
\newcommand{\fK}{f_\mathrm{K}}
\newcommand{\Tkin}{T_\mathrm{kin}}

\begin{document}

\title{Kinetic theory of cavity cooling and self-organisation of a cold gas}

\author{Wolfgang Niedenzu}
\affiliation{Institut f{\"u}r Theoretische Physik, Universit{\"a}t Innsbruck, Technikerstra{\ss}e~25, 6020~Innsbruck, Austria}
\author{Tobias Grie{\ss}er}
\affiliation{Institut f{\"u}r Theoretische Physik, Universit{\"a}t Innsbruck, Technikerstra{\ss}e~25, 6020~Innsbruck, Austria}
\author{Helmut Ritsch}
\email{Helmut.Ritsch@uibk.ac.at}
\affiliation{Institut f{\"u}r Theoretische Physik, Universit{\"a}t Innsbruck, Technikerstra{\ss}e~25, 6020~Innsbruck, Austria} 

\begin{abstract}
We study spatial self-organisation and dynamical phase-space compression of a dilute cold gas of laser-illuminated polarisable particles in an optical resonator. Deriving a non-linear Fokker--Planck equation for the particles' phase-space density allows us to treat arbitrarily large ensembles in the far-detuning limit and explicitly calculate friction forces, momentum diffusion and steady-state temperatures. In addition, we calculate the self-organisation threshold in a self-consistent analytic form. For a homogeneous ensemble below threshold the cooling rate for fixed laser power is largely independent of the particle number. Cooling leads to a $q$-Gaussian velocity distribution with a steady-state temperature determined by the cavity linewidth. Numerical simulations using large ensembles of particles confirm the analytical threshold condition for the appearance of an ordered state, where the particles are trapped in a periodic pattern and can be cooled to temperatures close to a single vibrational excitation.
\end{abstract}

\date{November 4, 2011}

\maketitle

\section{Introduction}
A dilute cold gas of polarisable particles can be manipulated in a controlled way using the light forces induced by a sufficiently strong laser far off any internal optical resonance~\cite{grimm2000}. For a free-space laser setup this force generates a conservative optical potential for the particles with a depth proportional to the local light intensity. As the forces are generated via photon redistribution among different spatial directions, the particles in turn alter the field distribution and act essentially as a spatially varying refractive index. While this backaction can safely be ignored in standard optical traps~\cite{miller1993}, it was shown to have a significant effect if the light fields are confined within an optical resonator enhancing the effective particle-light interaction~\cite{domokos03}. 
\par
For transverse illumination a threshold pump intensity where this coupled particle-field dynamics can lead to spatial selfordering of the particles into a regular pattern---resembling very closely a phase transition---was theoretically predicted and experimentally confirmed~\cite{domokos2002b,black2003,asboth05}. Due to cavity losses this dynamics is dissipative and thus can constitute a new cooling mechanism for a very general class of polarisable objects~\cite{vuletic2000,lev2008}. Extensive simulations using fairly large numbers of particles with large detunings predict that already with current molecular sources and cavity technology a useful phase-space compression could be achieved~\cite{salzburger2009}. However, particle-based simulations cannot be applied to sufficiently large particle numbers and laser powers for the whole parameter range of interest. As an alternative, a Vlasov-equation-based approach~\cite{griesser2010} for the particles' phase-space distribution provides a description for arbitrarily large ensembles, but excludes correlations important on longer time scales. In this work we adopt methods from plasma kinetic theory to generalise the Vlasov model to include correlations, leading to a non-linear Fokker--Planck equation for the statistically averaged phase-space distribution, which includes friction and diffusion and allows to predict cooling time scales and the unique steady-state distribution.

\section{Semiclassical equations of motion}
\begin{figure}
  \centering
  \includegraphics[width=0.69\columnwidth]{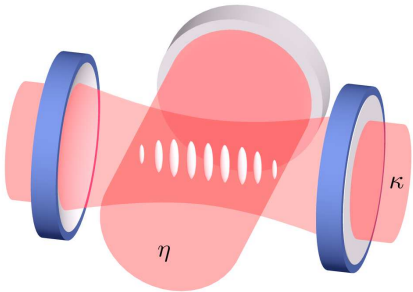}
  \caption{(Colour on-line) An ensemble of particles is illuminated by a transverse standing-wave laser and scatters light into the resonator (effective pump strength $\eta$). Above a threshold pump intensity the particles selforganise in a periodic pattern. Cavity losses are characterised by the decay rate $\kappa$.}\label{fig_resonator}
\end{figure}

We consider $N$ polarisable particles moving along the axis of a lossy standing-wave resonator assuming strong transversal confinement. The particles are off-resonantly illuminated by a transverse standing-wave pump laser and scatter light, whose phase is determined by the particle positions, into the cavity. The quantum master equation describing this system can be transformed into a partial differential equation for the Wigner function~\cite{domokos01}. Truncating this equation at second-order derivatives (semiclassical limit) yields a Fokker--Planck equation, which for positive Wigner functions---excluding all non-classical states~\cite{schleich}---is equivalent to the It\=o stochastic differential equation (SDE) system 
\begin{subequations}\label{eq_sde}
\begin{align}
\dd x_j &= v_j\,\dd t\\
\dd v_j &= -\frac{1}{m}\frac{\partial U(x_j,\alpha)}{\partial x_j}\,\dd t\\
\dd \alpha &= \left[\ii\left(\DC-U_0 \sum_{j=1}^N \sin^2(kx_j)\right)-\kappa\right]\alpha\,\dd t-\notag\\&\quad-\ii\eta\sum_{j=1}^N \sin(kx_j)\,\dd t+\sqrt{\frac{\kappa}{2}}\left(\dd W_1+\ii\,\dd W_2\right),\label{eq_sde_alpha}
\end{align}
\end{subequations}
with the single-particle potential
\begin{equation}
U(x,\alpha)=\hbar U_0|\alpha|^2 \sin^2(kx)+\hbar\eta (\alpha+\alpha^*) \sin(kx).
\end{equation}
Each particle has the mass $m$; $x_j$ and $v_j$ denote the centre-of-mass position and velocity of the $j$th particle, respectively. $U_0$ is the light shift per photon and $\kappa$ the cavity decay rate; the associated input noise is taken into account by the two Wiener processes $\dd W_1$ and $\dd W_2$. Here we have neglected momentum diffusion caused by spontaneous emission, which is valid for large ensembles and large detunings~\cite{lev2008}. The transverse laser standing wave gives an effective position-dependent pump of magnitude $\eta$ of the field mode $\alpha$ with wave number $k$. This laser is detuned by $\DC=\omega_\mathrm{p}-\omega_\mathrm{c}$ from the bare cavity resonance frequency. We will focus on the weak-coupling limit $N|U_0|\ll|\DC|\sim\kappa$ throughout this paper. Refer to fig.~\ref{fig_resonator} for a schematic view of the system.

\par
We used the scheme proposed in~\cite{kloeden77,ruemelin82} for the direct numerical integration of the SDE system~\eqref{eq_sde}. However, for analytical predictions and the description of very large ensembles, a continuous phase-space description as described below proves far more suitable. 

\section{Continuous description, instability threshold}
The semiclassical SDEs~\eqref{eq_sde} are equivalent to the Klimontovich equation~\cite{montgomery}
\begin{equation}\label{eq_klimontovich}
\frac{\partial \fK}{\partial t}+v\frac{\partial \fK}{\partial x}-\frac{1}{m}\frac{\partial U}{\partial x}\frac{\partial \fK}{\partial v}=0
\end{equation}
together with an evolution equation for the field mode $\alpha$ obtained replacing the sums in eq.~\eqref{eq_sde_alpha} by the integrals $N\iint\bullet \fK(x,v,t)\dd x\dd v$. $\fK(x,v,t)$ is the so-called Klimontovich or ``exact'' distribution function with initial condition
\begin{equation}
\fK(x,v,0)=\frac{1}{N}\sum_{j=1}^N\delta\big(x-x_j(0)\big)\delta\big(v-v_j(0)\big).
\end{equation}
As this function is highly irregular, the above reformulation has no computational merit by itself, but provides an ideal starting point for a statistical treatment. To this end we decompose every quantity ($\fK$, $\alpha$ and $U$) into its smooth mean value and fluctuations,
\begin{equation}
\fK(x,v,t)=:\ew{f(x,v,t)}+\delta f(x,v,t),
\end{equation}
with $\ew{\delta f(x,v,t)}=0$. The statistical average $\ew\bullet$ is over an ensemble of similar initial conditions $\{(x_j(0),v_j(0))\}$ and $\alpha(0)$, as well as over the realisations of the white noise process mimicking the input noise for the cavity field.
\par
For the smooth ensemble-averaged Klimontovich distribution, called \emph{one-particle distribution function}, we then find
\begin{equation}\label{eq_ensav}
\frac{\partial \ew{f}}{\partial t}+v\frac{\partial\ew{f}}{\partial x}-\frac{1}{m}\frac{\partial \ew{U}}{\partial x}\frac{\partial\ew{f}}{\partial v}=\ew{\frac{\partial\delta U}{\partial x}\frac{\partial \delta f}{\partial v}}.
\end{equation}
Neglecting all correlations in eq.~\eqref{eq_ensav} leads to the Vlasov equation, which becomes exact in the limit $N\rightarrow\infty$~\cite{griesser2010}. 
\par
The Vlasov equation together with the equation for $\ew{\alpha}$ possesses an infinite number of possible spatially homogeneous stationary solutions with zero cavity field, of which, however, not all are necessarily stable against perturbations. Indeed, for any (dimensionless) symmetric velocity distribution $g(v/\vT):=L\vT \ew{f(v)}$ and $\delta:=\DC-NU_0/2<0$ we find, that if
\begin{equation}\label{eq_critical_value}
\frac{N\eta^2}{\kB T}\vp\int_{-\infty}^\infty\frac{g^\prime(\xi)}{-2\xi}\dd \xi>\frac{\delta^2+\kappa^2}{\hbar|\delta|},
\end{equation}
where $\vp$ denotes the Cauchy principal value, perturbations trigger a self-organisation process. Here we have defined the thermal velocity $\vT^2:=2\kB T/m$; $L$ is the cavity length. For a Gaussian distribution the integral evaluates to one. The relation~\eqref{eq_critical_value} has been derived by methods presented in~\cite{griesser2010}. There it was shown that the threshold can be computed from the zeros of the dispersion relation ($\realt(s)>0$)
\begin{multline}\label{eq_dispersion}
D(s)=(s+\kappa)^2+\delta^2-\\-\ii\hbar k\delta\frac{NL\eta^2}{2m}\int_{-\infty}^\infty\left(\frac{\partial_v\ew{f}}{s+\ii kv}-\frac{\partial_v\ew{f}}{s-\ii kv}\right)\dd v.
\end{multline}

\section{Kinetic equation for the velocity distribution in the non-organised phase}
Let us now return to eq.~\eqref{eq_ensav}, which for weak spatial inhomogeneity---as expected below the instability threshold~\eqref{eq_critical_value}---approximately reads
\begin{equation}\label{eq_ensav2}
\frac{\partial \ew{f}}{\partial t}\approx\overline{\ew{\frac{\partial\delta U}{\partial x}\frac{\partial \delta f}{\partial v}}},
\end{equation}
where the overbar denotes the spatial average. The right-hand-side correlation function can be computed using established methods from plasma physics as in~\cite{montgomery} when the fluctuations evolve on a much faster time scale than the average values, which are considered ``frozen''. This is justified as long as the system remains far from instability. After some lengthy calculations, which we omit here for the sake of compactness, we obtain (for symmetric distributions) the non-linear Fokker--Planck equation
\begin{equation}\label{eq_nonlin_fpe}
\frac{\partial}{\partial t}\ew{F}+\frac{\partial}{\partial v}\Big(A[\ew{F}]\ew{F}\Big)=\frac{\partial}{\partial v}\left(B[\ew{F}]\frac{\partial}{\partial v}\ew{F}\right)
\end{equation}
for the velocity distribution $\ew{F(v,t)}:=L\ew{f(v,t)}$, with the coefficients
\begin{subequations}
\begin{align}
A[\ew F]&:=\frac{2\hbar k\delta\kappa \eta^2}{m}\frac{kv}{|D(\ii kv)|^2}\\
B[\ew F]&:=\frac{\hbar^2k^2\eta^2\kappa}{2m^2}\frac{\kappa^2+\delta^2+k^2v^2}{|D(\ii kv)|^2}.
\end{align}
\end{subequations}
This equation, which describes the sub-threshold gas dynamics, is one of the central analytical results of this work. Note that the coefficients functionally depend on $\ew{F}$ through the dispersion relation $D(\ii \omega):=\lim_{\varepsilon\downarrow 0}D(\varepsilon+\ii\omega)$. $A$ and $B$ represent the deterministic part of the equations and the field noise, respectively. All cavity-mediated long-range particle interactions are encoded in the dispersion relation. Note that very far below threshold the full dispersion relation~\eqref{eq_dispersion} reduces to $D(\ii kv)\simeq(\ii kv+\kappa)^2+\delta^2$, which corresponds to the $N$ independent-particles case. 

\section{Equilibrium distribution and temperature}
\begin{figure}
  \centering
  \includegraphics[width=8cm]{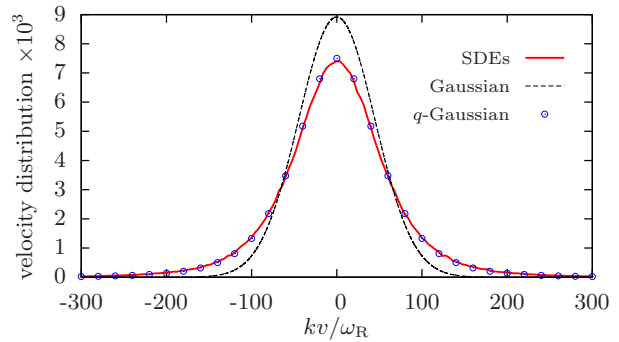}
  \caption{(Colour on-line) Normalised $q$-Gaussian velocity distribution for $\delta=-2.5\omrec$ obtained by integrating the semiclassical equations~\eqref{eq_sde} up to $\omrec t=5N$ (solid line) compared to the theoretically predicted $q$-Gaussian~\eqref{eq_qgauss} (circles). The Gaussian (dashed line) is plotted for the temperature~\eqref{eq_temperature}. The kinetic temperature differs considerably from eq.~\eqref{eq_temperature}, $\kB \Tkin=2.5\kB T$. Parameters: $N=5000$, $NU_0=-0.1\omrec$, $\sqrt{N}\eta=1800\omrec$, $\kappa=100\omrec$ and $\DC=-2.55\omrec$. Ensemble average over $25$ initial conditions and $10$ realisations of the white noise process.}\label{fig_qgauss}
\end{figure}

Normalisable steady-state solutions of eq.~\eqref{eq_nonlin_fpe} exist only for negative detuning $\delta<0$, where light scattering is accompanied by kinetic-energy extraction from the motion. For positive detuning the particles are heated. Interestingly, we obtain the non-thermal $q$-Gaussian velocity distribution function (also known as Tsallis distribution and closely related to Student's $t$-distribution)~\cite{deSouza1997,lutz2003}
\begin{equation}\label{eq_qgauss}
\ew{F(v)}\propto\left(1-(1-q)\frac{mv^2}{2\kB T}\right)^\frac{1}{1-q},
\end{equation}
with $q:=1+\omrec/|\delta|$ and the recoil frequency $\omrec:=\hbar k^2/2m$. We have defined an effective ``temperature'' parameter
\begin{equation}\label{eq_temperature}
\kB T=\hbar\frac{\kappa^2+\delta^2}{4|\delta|}=\frac{\hbar\kappa}{2}.
\end{equation}
The latter minimum value of $T$ appears for $\delta=-\kappa$. The parameter $|\delta|/\omrec$ determines the shape of the distribution and gives rise to further restrictions. Normalisable solutions exist for $|\delta|>\omrec/2$, the second moment (kinetic energy) only for $|\delta|>3\omrec/2$. The case $|\delta|=\omrec$ corresponds to a Lorentzian distribution. For $|\delta|/\omrec\rightarrow\infty$, \ie $q\rightarrow 1$, the distribution~\eqref{eq_qgauss} converges to a Gaussian distribution with kinetic temperature $\kB \Tkin := m\ew{v^2}$ given by the parameter $\kB T$ defined in eq.~\eqref{eq_temperature}, which justifies the choice of ``temperature'' above. Refer to fig.~\ref{fig_qgauss} for an example of prominent $q$-Gaussian behaviour ($q=1.4$). Tsallis distributions have already been observed experimentally in dissipative optical lattices~\cite{douglas2006}. 

\par
Of course the distribution~\eqref{eq_qgauss} can be unstable and in fact is, if
\begin{equation}\label{eq_selfconsistent_threshold}
\sqrt{N}\eta>\sqrt{N}\etac:=\frac{\kappa^2+\delta^2}{2|\delta|}\sqrt{\frac{2}{3-q}}\stackrel{\delta=-\kappa}{=}\kappa\sqrt{\frac{2}{3-q}},
\end{equation}
which can be seen by inserting the $q$-Gaussian into~\eqref{eq_critical_value}. For a Gaussian and $\delta=-\kappa$ this criterion may be reformulated as
\begin{equation}
N|U_0|V_\mathrm{opt}>\hbar\kappa^2,
\end{equation}
where $V_\mathrm{opt}$ is the optical potential depth created by the pump laser. If the condition~\eqref{eq_selfconsistent_threshold} is satisfied, we conclude that there exists no spatially homogeneous steady state at all and consequently we expect the system to organise. The possibility of attaining such an inhomogeneous equilibrium even though the uniform distribution~\eqref{eq_qgauss} is stable cannot be ruled out mathematically but seems unlikely on physical grounds. These expectations are confirmed by all our numerical simulations of eqs.~\eqref{eq_sde}. 
\par
Based on these considerations we predict the occurrence of dissipation-induced self-organisation if the self-consistent relation~\eqref{eq_selfconsistent_threshold} is fulfilled. That is, any initially stable, unorganised distribution will loose kinetic energy (cavity cooling) and eventually selforganise. Contrary to the self-organisation process of an initially unstable state, which is abrupt and accompanied by strong heating, dissipation-induced self-organisation is characterised by a much slower buildup of the photon number and monotonous cooling.  

\section{Selforganised phase}

Above threshold the mathematics becomes more complex due to the inhomogeneous spatial distribution. However, using action-angle variables~\cite{luciani1987,chavanis2007}, we can still derive a Fokker--Planck equation similar to eq.~\eqref{eq_nonlin_fpe}~\cite{griesser2011selforganisation}. In the limit of deep trapping, where a harmonic approximation for the potential becomes valid, the steady-state solution for the strongly organised phase is a thermal distribution with a kinetic temperature depending on the linewidth $\kappa$ and the trap frequency $\omega_0$,
\begin{equation}\label{eq_temperature_so}
\kB \Tkin = \hbar\frac{\kappa^2+\delta^2+4\omega_0^2}{4|\delta|}.
\end{equation}
The trap frequency is $\omega_0^2=4\eta\omrec\ew{|\realt\alpha_\infty|}$ and can be approximated by
\begin{equation}
\omega_0^2\simeq\sqrt{N}\eta\omrec\left(\frac{\eta}{\etac}+\sqrt{\frac{\eta^2}{\etac^2}-1}\right)\sim N
\end{equation}
in the far-detuned regime where $|\delta|\gg\omrec$; $\etac$ is the self-consistent critical value defined in eq.~\eqref{eq_selfconsistent_threshold}. As the temperature depends explicitly on the laser power, higher pump strengths result in deeper trapped ensembles with increased kinetic energy. Note that this system has the interesting property that the more particles we add, the deeper the optical potential gets as is the case for self-gravitating systems~\cite{posch2005}.
\par

\begin{figure}
  \centering
  \includegraphics[width=8cm]{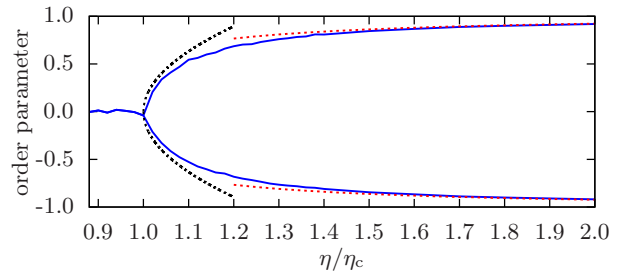}
  \caption{(Colour on-line) Order parameter~\eqref{eq_orderparameter} as function of the pump strength obtained from eqs.~\eqref{eq_sde} in the long-time limit ($\omrec t=50N$). The red dashed line is given by eq.~\eqref{eq_theta_largeeta}, the black dashed line around the critical point corresponds to eq.~\eqref{eq_theta_smalleta}. Parameters: $N=1000$, $NU_0=-\omrec$, $\kappa=100\omrec$, $\DC=NU_0/2-\kappa$ (\ie $\delta=-\kappa$) and $\kB T_0=110\Erec$. Ensemble average of $20$ (away of the critical point) and $60$ (around the critical point) noise trajectories, respectively, for one initial condition.}\label{fig_orderparameter}
\end{figure}

The order parameter
\begin{equation}\label{eq_orderparameter}
\Theta:=\lim_{t\rightarrow\infty}\iint\sin(kx)\ew{f(x,v,t)}\dd x\dd v  
\end{equation}
is an adequate measure of particle localisation in equilibrium as it is zero for a completely homogeneous distribution and plus/minus one for the perfectly selforganised phase (\ie $\delta$-peaks), the sign depends on whether the odd- or even wells are populated~\cite{asboth05}. In fig~\ref{fig_orderparameter} we depict its behaviour as a function of the pump strength. Initially, the particles were spatially homogeneously distributed with a (stable) Gaussian velocity distribution; self-organisation sets in because of the cavity cooling effect. Hence the branching point is given by the self-consistent threshold~\eqref{eq_selfconsistent_threshold}. 

\par
For the strongly organised phase, where the distribution function is a thermal state with temperature~\eqref{eq_temperature_so}, the order parameter reads (in harmonic oscillator approximation)
\begin{equation}\label{eq_theta_largeeta}
\Theta=\pm\left(1-\frac{\kB \Tkin}{\hbar\omega_0^2}\omrec\right).
\end{equation}
Its maximum value is limited by the detuning and the recoil frequency, $\Theta\rightarrow\pm\left(1-{\omrec}/{|\delta|}\right)$ for $\eta\rightarrow\infty$. This corresponds to a Gaussian spatial distribution with width $k^2(\Delta x)^2=2{\omrec}/{|\delta|}$. Let us also briefly investigate the opposite limit of pump strengths slightly above threshold. Solving the self-consistency equation
\begin{equation}
\realt\ew\alpha_\infty=\frac{-N\eta|\delta|}{\kappa^2+\delta^2}\Theta
\end{equation}
perturbatively around the critical point for a thermal state yields
\begin{equation}\label{eq_theta_smalleta}
\Theta\simeq \pm 2 \sqrt{\frac{\eta}{\etac}-1}
\end{equation}
and thus a critical exponent of $1/2$, as already predicted in~\cite{asboth05}.
\par
\begin{figure}
  \centering
  \includegraphics[width=8cm]{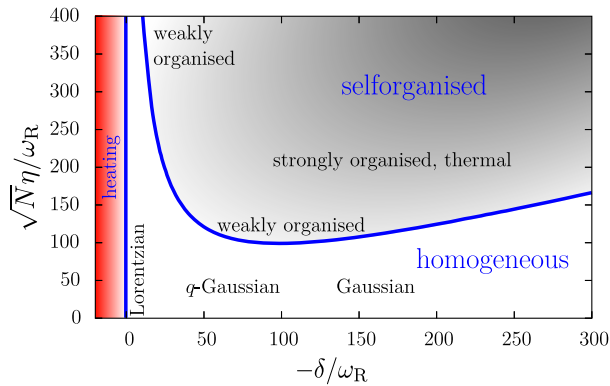}
  \caption{(Colour on-line) Schematic view of the phase diagram in the weak-coupling limit ($N|U_0|\ll\kappa$) for $\kappa=100\omrec$. Equilibrium solutions exist only for $\delta<-\omrec/2$, the Lorentzian corresponds to the case $|\delta|=\omrec$. For large values of the detuning $|\delta|$, strongly organised equilibria exist already for pump strengths slightly above the critical value, \cf also fig.~\eqref{fig_orderparameter}.}\label{fig_phasediagram}
\end{figure}

The self-consistent phase diagram including cavity cooling is sketched in fig.~\ref{fig_phasediagram}. For small $|\delta|/\omrec$ the ensemble will always remain weakly organised (\ie a small spatial modulation on top of a homogeneous background) above threshold as the necessary prerequisites for a strongly organised equilibrium---and hence the validity of eq.~\eqref{eq_temperature_so}---cannot be fulfilled. 

\section{Cooling time}
\begin{figure}
  \centering
  \includegraphics[width=8cm]{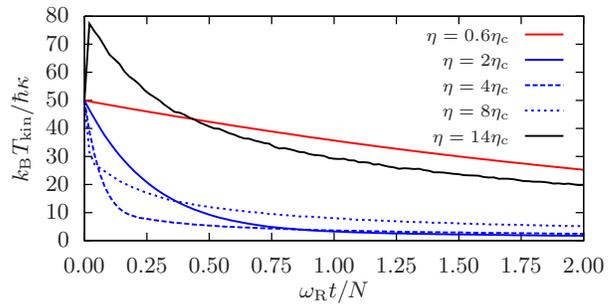}
  \caption{(Colour on-line) Kinetic temperature for $N=500$ and different pump strengths. Parameters: $\kappa=100\omrec$ and $\delta=-\kappa$. The initial instability threshold is at $\eta=10\etac$. Ensemble average over $5$ initial conditions with $5$ noise trajectories.}\label{fig_fixedN_variableeta}
\end{figure}

Let us take a closer look at the cooling time $\tau$, \ie the time characteristic for the kinetic-energy equilibrisation. First we treat the case of a fixed ensemble size and variable pump strength. The drift term $A+\partial_vB$ in the non-linear Fokker--Planck equation~\eqref{eq_nonlin_fpe}---scaling as $\sim\eta^2/\omrec$---might suggest the conclusion that the larger the pump strength the shorter the cooling time. However, numerical simulations (\cf fig.~\ref{fig_fixedN_variableeta}) prove this expectation to be somewhat misleading. The reason therefor is the onset of self-organisation which occurs as soon as the momentary distribution becomes unstable according to eq.~\eqref{eq_critical_value}. Particle trapping is a hindrance to optimal cooling in a twofold way. Firstly, with its appearance kinetic energy is dissipated at a significantly lower rate because a part of the laser power is utilised for the buildup of the potential and consequently is not available for cooling. Secondly, the lowest achievable temperature increases with the pump strength, \cf eq.~\eqref{eq_temperature_so}. The situation is worst for initially unstable ensembles since the particles are heated before being cooled.  Hence it is a plausible requirement that self-organisation has to be avoided to realise the optimal cooling time. Accordingly, as a rule of thumb we may state that the latter is achieved for a laser power which renders the desired gas temperature critical. Hence the optimal cooling time is estimated from eq.~\eqref{eq_nonlin_fpe} to be
\begin{equation}
\tau_\mathrm{opt}\approx\frac{kv_{\mathrm{T}_0}}{4\sqrt{\pi}\kappa^2}N
\end{equation}
assuming a Gaussian and $\delta=-\kappa$ for simplicity. This estimate is valid for $kv_{\mathrm{T}_0}\gg\kappa$, where $T_0$ is the initial temperature.
\par
\begin{figure}
  \centering
  \includegraphics[width=8cm]{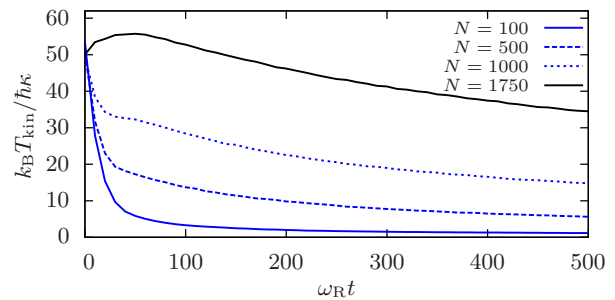}
  \caption{(Colour on-line) Kinetic temperature for fixed pump strength and $N=\{100,500,1000,1750\}$, from bottom to top.. Parameters: $\eta=28\omrec$, $\kappa=100\omrec$ and $\delta=-\kappa$. The self-consistent threshold~\eqref{eq_selfconsistent_threshold} is surpassed for all curves; the ensemble with $N=1750$ is initially unstable. Ensemble average over $25$ ($N=100$) and $8$ (higher particle numbers) initial conditions, respectively, and $5$ noise trajectories.}\label{fig_fixedeta_variableN}
\end{figure}

For a fixed $\eta$, all ensembles composed of $N<N_\mathrm{c}$ particles experience in a good approximation the same cooling time $\tau\sim \omrec/\eta^2$ for reaching the minimal temperature~\eqref{eq_temperature}. $N_\mathrm{c}$ is the critical particle number rendering the given $\eta$ critical. Note however, that this time scale is suboptimal for all ensemble sizes except $N=N_\mathrm{c}$. The scaling with $N$ for fixed laser power is more favourable as for standard cavity cooling~\cite{horak01b}. Refer to fig.~\ref{fig_fixedeta_variableN}. For initially stable ensembles the cooling rate is approximately the same until the instability point is reached. 

\par
\begin{figure}
  \centering
  \includegraphics[width=8cm]{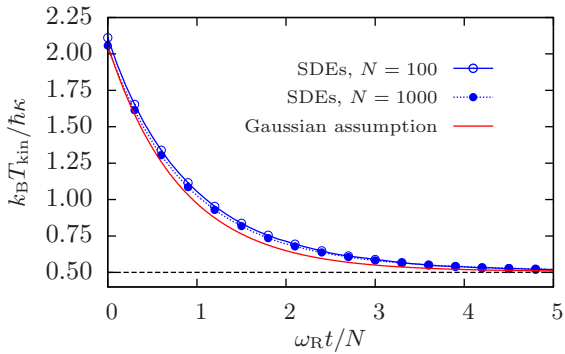}
  \caption{(Colour on-line) Comparison of the kinetic temperature obtained from the SDEs~\eqref{eq_sde} and as solution of eq.~\eqref{eq_temperature_evolution} for two different particle numbers and constant $\sqrt{N}\eta$. As expected, both ensembles evolve on a time scale $\sim N/\omrec$. Parameters: $NU_0=-0.01\omrec$, $\kappa=100\omrec$, $\DC=NU_0/2-\kappa$ (\ie $\delta=-\kappa$) and $\sqrt{N}\eta=80\omrec\equiv0.8\sqrt{N}\etac$. Ensemble average of $50$ ($N=100$) and $2$ ($N=1000$) initial conditions, respectively, and $50$ realisations of the white noise process.}\label{fig_temperature}
\end{figure}
\begin{figure}
  \centering
  \includegraphics[width=8.5cm]{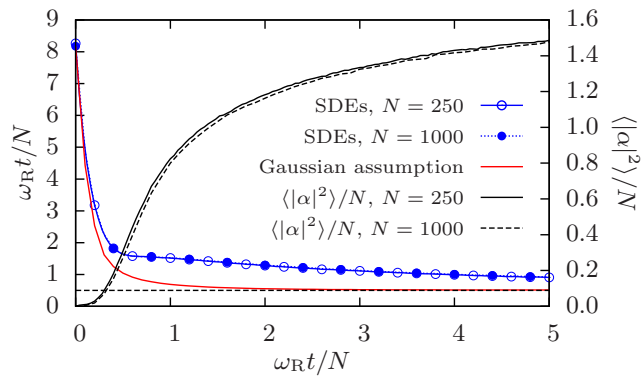}
  \caption{(Colour on-line) Temperature evolution above the self-consistent threshold~\eqref{eq_selfconsistent_threshold}. The homogeneous distribution is stable until the relation~\eqref{eq_critical_value} is satisfied. This also limits the validity of eq.~\eqref{eq_temperature_evolution}. Parameters: $NU_0=-\omrec$, $\kappa=100\omrec$, $\DC=NU_0/2-\kappa$ (\ie $\delta=-\kappa$) and $\sqrt{N}\eta=200\omrec\equiv 2\sqrt{N}\etac$. As $\sqrt N\eta=\text{const.}$, the photon number scales only $\sim N$ and not $\sim N^2$ (superradiance effect). Furthermore, the equilibrium temperature is the same for both ensembles. The simulations were performed up to $\omrec t=40N$ and revealed a temperature $\kB T_\mathrm{kin}\approx 0.57\kappa$, which agrees very well with the theoretical prediction~\eqref{eq_temperature_so}. Ensemble average of $50$ ($N=250$) and $25$ ($N=1000$) initial conditions, respectively, and $10$ realisations of the white noise process. }\label{fig_above}
\end{figure}

Let us briefly mention another aspect of the scaling of the Fokker--Planck equation~\eqref{eq_nonlin_fpe} particularly useful for numerical simulations. Fixing $\sqrt{N}\eta$ for different particle numbers yields a cooling time $\tau\sim N/\omrec$. Numerical simulations of the semiclassical equations~\eqref{eq_sde} confirm this result, \cf figs.~\ref{fig_temperature} and~\ref{fig_above}. There we have depicted the temperature evolution for the pump strength being a fixed fraction of the critical value $\etac$ for two different particle numbers, \ie $\sqrt{N_1}\eta_1=\sqrt{N_2}\eta_2$. In fig.~\ref{fig_above} the threshold~\eqref{eq_critical_value} is surpassed during the time evolution.

\par
In order to verify the Fokker--Planck equation~\eqref{eq_nonlin_fpe}, we consider the equation
\begin{equation}\label{eq_temperature_evolution}
\frac{\dd}{\dd t}\kB \Tkin =-2m\int_{-\infty}^\infty v\left(-A\ew{F}+B\frac{\partial}{\partial v}\ew{F}\right)\dd v.
\end{equation}
for the kinetic temperature $\kB \Tkin:=m\ew{v^2}$ and close it by assuming $\ew{F(v,t)}$ to be a Gaussian velocity distribution with temperature $\kB \Tkin$ for all times. This assumption is well justified for large detunings $|\delta|\gg\omrec$ and its predicted temperature reproduces the results obtained from the SDEs~\eqref{eq_sde} quite well, \cf fig.~\ref{fig_temperature}. The main difference between the curves stems from the relatively small particle numbers in combination with pump values close to threshold, where the hypothesis of separated time scales used to derive eq.~\eqref{eq_nonlin_fpe} is no longer valid due to long-lived fluctuations.
\par
Of course, a thorough investigation of the validity of eq.~\eqref{eq_nonlin_fpe} would require a numerical integration thereof. However, in steady state, all results of numerical simulations of the SDE system~\eqref{eq_sde} were found to be in excellent agreement with the predictions of the kinetic theory, both, below (\eg fig.~\ref{fig_qgauss}) and above (\eg fig.~\ref{fig_above}) threshold.
\\\\
\section{Conclusion and outlook}
Collective light scattering from a dilute gas of cold particles into a high-$Q$ resonator mode under suitable conditions leads to friction forces and cooling of particle motion even below the self-organisation threshold. In contrast to standard cavity cooling the friction force below threshold only weakly depends on the particle number and leads to fast internal thermalisation towards a $q$-Gaussian velocity distribution of particles with average kinetic energy determined by the cavity linewidth only. Thus this constitutes a viable method for cooling very large ensembles of particles with sufficient polarisability independent of the need for a cyclic optical transition as well as a way to implement evaporative cooling for low densities, where no direct collisions occur.
\par
As only the polarisability and mass of a particle enter, one can easily envisage a combination of many species within the cavity to commonly interact with the same pump and cavity fields as a generalised form of sympathetic cooling without the need of direct interparticle interactions. The corresponding non-linear equation replacing the Fokker--Planck equation~\eqref{eq_nonlin_fpe} will then be of Balescu--Lenard type, cf.~\cite{griesser2011selforganisation} and references therein.

\begin{acknowledgments}
This work has been supported by the Austrian Science Fund FWF through projects P20391 and F4013. We would like to thank J\'anos Asb\'oth, Hessam Habibian, Giovanna Morigi and Matthias Sonnleitner for helpful discussions.
\end{acknowledgments}


\end{document}